# Equitable Optimization of U.S. Airline Route Networks

Andy G. Eskenazi[*], Arnav P. Joshi[‡], Landon G. Butler[+], Megan S. Ryerson[◊]

[*]Department of Mechanical Engineering and Applied Mechanics, University of Pennsylvania, Philadelphia, PA

[*]Department of Mathematics, University of Pennsylvania, Philadelphia, PA

[+, ‡, ◊]Department of Electrical and Systems Engineering, University of Pennsylvania, Philadelphia, PA

[◊]Department of City and Regional Planning, University of Pennsylvania, Philadelphia, PA

*Email:* andyeske@sas.upenn.edu[*], landonb3@seas.upenn.edu[+], arnavj@seas.upenn.edu[‡], mryerson@design.upenn.edu[◊]

*Abstract*—Restructuring route networks (i.e., modifying the graph of origin-destination pairs) remains a promising alternative for reducing the airline industry's environmental impact. However, there exists a fundamental trade-off between emissions from flight and airport accessibility, since flights connecting underserved, low-accessibility communities tend to possess high $CO_2$/seat-mile ratios. Thus, this work develops an open-source analytical framework and methodology that restructures U.S. airline route networks to simultaneously minimize emissions and maximize airport accessibility. To achieve this goal, this paper designs a metric to quantify airport accessibility and combines it with an open-source system-wide emissions estimation methodology. This facilitates the creation of a mixed-integer linear optimization model that returns revised flight frequencies and aircraft allotment. Using United Airlines 2019 Q3 data as a case study, this model is able to construct an alternative route network with a 25% reduction on the total number of flights, 4.4% decrease in the average emissions per seat-mile and a 17.6% improvement in the spread of the airports' accessibility scores, all while satisfying historic passenger demand.

## 1. INTRODUCTION

The airline industry is one of the foremost contributors to the global emissions of carbon dioxide ($CO_2$). In 2019 alone, the sector released 915 million tons of $CO_2$ (ETC, 2021) into the atmosphere, an amount equal to 2.5% of the global emissions. Considering, as well, the production of other harmful greenhouse gases like carbon monoxide (CO), nitrogen oxides ($NO_x$) and hydrocrabons (HC), in addition to the formation of contrail clouds, the airline industry accounts for 3.5% of the total anthropogenic global warming (Lee and Forster, 2021). The emissions generated by the sector are estimated to continue growing at an average annual rate of 3.3% (Graver et al., 2019); over the next 30 years, this growth could release an additional 40 Gt of $CO_2$, resulting in a situation that could exhaust anywhere between 10% (ETC, 2021) to 25% (Yeo and Pidcock, 2016) of the carbon budget designed to limit the global increase in temperature outlined by the Paris Climate Agreement.

Decarbonizing aviation is paramount. Accomplishing this mission will require significant investments in green propulsion technologies driven by fuel cells (Wright, 2022) and batteries (Lampert, 2021), as well as new propellants such as hydrogen (Airbus, 2020) and sustainable aviation fuel (Muller and King, 2021). However, these concepts are still on their early development phases, and it may take decades before their testing is completed and deployment takes place (IATA, 2019). However, planning for the decarbonization of civil aviation cannot be performed over uncertainties, given the industry's massive environmental impact and the urgency that reducing greenhouse emissions signifies for human subsistence

One approach that would not require the development of new technologies and could see an immediate reduction in the industry's $CO_2$ emissions is the restructuring of airline route systems (Scheelhaase et al., 2010; Wen, 2013). Through the careful selection of routes and flight frequencies, as well as aircraft allotment, it is possible to redesign airline route networks (i.e., redraw the graph of origin-destination pairs) to reduce the system's cumulative greenhouse gas emissions while satisfying the current level of passenger demand.

Previous research on the green restructuring of airline route networks has sought to examine the environmental impact of hub-and-spoke against point-to-point models in Europe (Morrell and Lu, 2007; Peeters, 2001), minimize the monetary cost of the European Union's Emissions Trading Scheme (ETS) charges (Wen, 2013) as well as reduce the locational climate impact due to concentrated air traffic through individual trajectory optimizations (Rosenow et al., 2017). While innovative and insightful, these studies introduced models that focused on a limited selection of aircraft-engine types (i.e., aircrafts equipped with a particular engine type), and thus were not able to fully estimate the environmental footprint of airlines with varied fleets. More critically, these models did not address the intrinsic trade-off between flight emissions and passenger access to airports (and thus to the greater air transportation system); as previous research has shown, short-haul, regional routes have the highest ratio of $CO_2$ emissions per seat-mile (Arul, S. G., 2015; Filippone and Parkes, 2021; Montaur et al., 2021, Eskenazi et al., 2022). Thus, from an emissions reduction perspective, these flights should be removed from a network. However, these routes are also essential in connecting remote communities to the greater air transportation system, especially when no other feasible mode of transportation is available (Fageda et al., 2018). Hence, the restructuring of airline route networks to minimize its environmental impact cannot come at the expense of reduced airport accessibility.

The contribution of this research lies in the provision of an analytical framework and methodology to realistically restructure airline route networks in the United States, with the dual goal of reducing emissions while maintaining or improving the level of airport accessibility, all while satisfying the historic passenger demand. This work develops an accessibility metric that leverages open-source data from the American Community Survey (ACS) to quantify and illustrate air transportation accessibility in each airport and census tract in the U.S. This metric is integrated with a stage-of-flight, aircraft and engine specific emissions estimation methodology that computes the $CO_2$ equivalent ($CO_2e$) emissions of four types of greenhouse gases ($CO_2$, HC, $NO_x$ and CO) for every operable flight between any two city pairs in the U.S., using publicly available data from the Bureau of Transportation Statistics (BTS), the Federal Aviation Administration (FAA), the International Civil Aviation Organization (ICAO) and EUROCONTROL. Together, both tools enable the construction of a novel optimization algorithm that takes as inputs the historic origin-destination (OD) passenger demand of a particular U.S. airline, alongside aircraft-engine specific flight emissions estimations and airport accessibility scores, and returns as outputs the restructured route system for that airline, with new flight frequencies, aircraft allotment and satisfied demand along each route. Furthermore, due to the formulation of the optimization's objective function, this work's analytical framework permits simulating various costs of $CO_2$ emissions and economic benefits of having more passengers flying, which in turn enables modeling future scenarios that can motivate policy and network planning decisions.

The remainder of this work is organized in the following way: a literature review of the emissions estimation methodology and passenger accessibility is presented in Section 2; an overview of the analytical framework and mathematical model is provided in Section 3; the inputs of the optimization are presented in Section 4; the route network restructuring algorithm, alongside the set of constraints, is detailed in Section 5; the results and analysis from implementing the methods are featured in Section 6, using United Airlines as a case study; and final conclusions are delivered in Section 7.

## 2. Literature Review

To construct a model that jointly reduced flight emissions while maximizing airport accessibility, it was necessary to first develop the tools to estimate both of these two quantities. The authors constructed an emissions estimation methodology detailed in Eskenazi et al. (2022) that leveraged publicly available datasets and could be implemented in a system-wide manner. Hence, the following literature review places a greater emphasis on accessibility metrics and the evolution of air service distribution within the United States since the Airline Deregulation Act of 1978. For a more comprehensive review of emissions estimations, the present study refers the interested reader to Eskenazi et al. (2022).

### 2.1. Flight Emissions

The literature pertaining to aviation emissions estimation methodologies is diverse and can be organized in the following categories. To start, some of the existing methodologies (Jardine, 2005; Jardine, 2009) can be used to calculate system-wide emissions (that is, of an entire airline route network), but do not take into consideration granular details such as the airtime in each stage of flight (e.g., landing, take-off, cruising), the aircraft type (e.g., A320), and the engine type (e.g., the CFM56-5B as opposed to the IAE V2500-A5, both of which can power an A320). Methodologies that do take into account these granular details are often restricted in their scope to a particular flight, airport, or geographical region (Tokuslu, 2021; Xu et al., 2020). While methodologies that can compute system-wide emissions using granular estimation approaches do exist, they are inaccessible to the academic community due to employing datasets that are often proprietary (Kim et al., 2007; Miyoshi and Mason, 2009).

Balancing the above-mentioned trade-offs, the authors of the present study proposed an open-source, data-driven emissions estimation methodology that aimed to address this gap in the literature. Their approach used six publicly available datasets, sourced from the United States' Bureau of Transportation Statistics (BTS), the Federal Aviation Administration (FAA), the International Civil Aviation Organization (ICAO) and EUROCONTROL. To show the methodology's system-wide capabilities, the authors computed the emissions associated for nearly 1.65 million scheduled domestic flights operated during 2019 Q3 in the United States.

Due to the current lack of open-source data, this emissions estimation methodology did not incorporate the impact of contrail formation, aircraft take-off weights, altitude variations along flight path, and weather patterns. Nonetheless, it still managed to produce granular results when compared to those proposed by the United Nations Environmental Program (Peeters and Williams, 2012). In particular, it returned varying emissions estimates for the same distance flown as it took into account the nuances of time flown, stages of flight, aircraft type and engine type. As a result, due to its open-source nature, relatively high accuracy and quick computational speed, in addition to its compatibility with publicly available data sets, the authors selected this methodology to estimate the emissions associated with each flight in the optimized airline route network. The data and the code used to calculate emissions can be found in an online repository (Eskenazi et al., 2022). A brief explanation of the methodology is provided in section 4.3. of the present study.

*2.2. Air Transportation Accessibility*

Since the enactment of the Airline Deregulation Act in 1978, airlines have focused on serving medium and large hub airports that possessed sufficient passenger demand to be economically sustainable. As Wei and Grubesic (2015) argue in their work, deregulation started a trend of service suspensions to regional, smaller airports due to higher costs, shifting the flight network distribution towards Hub-and-Spoke, which is more efficient from an operational standpoint (Jenkins, 2011). Between 1980 and 2010, hubbing has generated a greater reliance on connecting rather than on non-stop traffic, reducing the total number of destinations with direct itineraries. This situation has slowly led to an unequal distribution of passenger accessibility in the United States, especially in the southeast and northwest (Jenkins, 2011). In fact, 29 out of 300 commercial airports dominate the majority of the air traffic in the country (Jenkins, 2011).

Thus, there have been various studies attempting to capture the nuances of air transportation accessibility in the United States, or the extent to which the country's air transportation system allows individuals to reach their destination, as Geurs and Van Wee (2004) define it. However, accessibility can take various forms depending on what aspect of it is being measured (locational, individual or economical), explaining why it is difficult to create an all-encompassing metric (Geurs and Van Wee, 2004). For example, locational accessibility is often employed to quantify proximities of communities to airports, the main point of entry of individuals to the air transportation system (Geurs and Van Wee, 2004), while potential accessibility is more suited to measure the opportunities of a particular region compared to all others.

These different forms of accessibility are reflected in the existing literature, such as in the research conducted by Wei and Grubesic (2015), which uses centrality and shortest path measures to quantify the geographical importance of airports in the U.S. Their results show that the more central an airport is, the lower its cumulative distance to all the other nodes. A similar study, produced by Matisziw and Grubesic (2010), defines a bimodal metric to capture the locational access to the air transportation system. In their work, the authors devise a model that incorporates both the costs of reaching a certain airport for every census tract polygon in the United States and the level of accessibility within the air transportation network. Using a distance threshold of 160 miles from every census tract, Matisziw and Grubesic (2010) evaluate the options that fall within that driving distance and conclude that there exist pronounced variations in locational accessibility among airports in the United States. Other studies, such as Reynolds-Feighan and McLay's (2006), define the accessibility metric by focusing on capacity and thus considers the total number of seats offered at a particular airport. The authors also extend their metric to examine the total number of seat miles available at each airport in the network, in an attempt to capture the reach of the destinations served. However, Reynolds-Feighan and McLay's (2006) research only considers direct flights and thus excludes connecting traffic from the analysis.

Despite differences in their approaches, all these metrics (Wei and Grubesic, 2015; Matisziw and Grubesic, 2010; Reynolds-Feighan and McLay, 2006) manage to effectively elucidate the fact that accessibility in the United States' air transportation system is unequally distributed. Recognizing the importance of this issue, the federal government has created the essential air services program (EAS). Established in 1978, EAS was born to ensure the connectivity of remote communities across the country against the backdrop of the airline deregulation act. Today, EAS is run by the US Department of Transportation (DoT), and the program subsidizes a number of carriers to serve 60 communities in Alaska and 115 in the lower 48 mainland states (U.S. Department of Transportation, 2017). However, in order for a carrier to receive an EAS subsidy to serve a certain airport, the DoT has established a number of criteria that must be met. Succinctly, the carrier must be willing to offer an average of 10 rotations per week to an airport that is located 175 miles or more from the nearest hub. In addition, the subsidy offered per passenger must be equal to $200 or less.

EAS has played a paramount role in ensuring that air services are provided to remote communities in the United States. However, since its inception, the program has been the object of numerous criticisms, the biggest being that service utilization has been steadily low (Grubesic and Wei, 2013). As a matter of fact, during 2010, more than 660,000 passengers were transported to and from EAS communities on about 103,000 flights, yielding an average of 6 passengers per rotation (Grubesic and Wei, 2013). Consequently, because these flights only carry a small number of passengers with a low-load factor over mostly short distances (Matisziw and Grubesic, 2012), their $CO_2$ emitted per seat-mile is quite high (Eskenazi et al., 2022). Furthermore, research has demonstrated that EAS flights are not the most efficient in actually increasing travel options for the populations they serve. As evidenced by the geographically informed model developed by Grubesic and Wei (2012), 85% of all EAS flights had an efficiency score below 80%, and in many cases, several EAS routes were redundant with existing, unsubsidized ones (Grubesic et al., 2012), leaving significant room for improvement.

In short, EAS flights do not fully accomplish their goal of integrating remote communities to the greater air transportation system of the United States. To correct this problem and enhance air transportation accessibility in the country, Matisziw and Grubesic (2012) suggest that a more effective means would be to directly reconfigure airline route networks and flight options. The challenge, however, is being able to achieve greater connectivity and integration while also decreasing the environmental impact stemming from flights themselves. This work seeks to precisely address this challenge: reconstructing airline route networks to simultaneously balance the intrinsic trade-off between flight emissions and accessibility. The following four sections proceed to elaborate on the implemented methodology and optimization algorithm.

## 3. MODEL OVERVIEW

In general terms, the optimization model proposed in this work was composed of three main parts, as can be in the **Fig. 1** schematic. First, the model inputs (the groundwork for the optimization), which consisted of the accessibility scores for every airport, the historic demand along each flown flight leg (origin-destination pair), as well as the aircraft-engine specific flight emissions for every potential flight leg, all corresponding to a particular U.S. airline during the study period. Next, the mathematical framework, which proposed utilizing a mixed-integer linear optimization model. And finally, the model outputs, which included the restructured route system for the airline, complete with new flight frequencies, aircraft allotment and satisfied demand along each route. In the present study, United Airlines' (UA) 2019 Q3 data (corresponding to July, August and September) was used as a case study to illustrate the optimization model's capabilities. The following three sections effectively expand on the model's three main components in greater detail.

## 4. MODEL INPUTS

### 4.1. Historic Origin-Destination Demand

A key component of the optimization model required calculating the historic origin-destination (OD) demand of all possible flight legs that could be operated by a single airline, regardless of whether the traffic was direct or connecting. To achieve this task, the present study utilized both the BTS' Airline Origin and Destination Survey (DB1B) Coupon (BTS, 2019) and Ticket (BTS, 2019) databases. The former permitted obtaining origin-destination (OD) data for airline tickets sold since 1993, classified by operating carrier and quarter of the year. The latter facilitated classifying itineraries between one-way and round-trip. In 2019 Q3, of the 528,974 itineraries operated by UA, it was found that 58.34% of the itineraries were round-trip, while the remaining 41.67% were one-way, as summarized in Table 1.

Finding the historic OD demand for one-way itineraries was relatively simple, given that the origin and destination airports could be taken as the first and last airports of a sequence of flights, and then the associated number of passengers flying on that itinerary would be assigned. In essence, a passenger on the

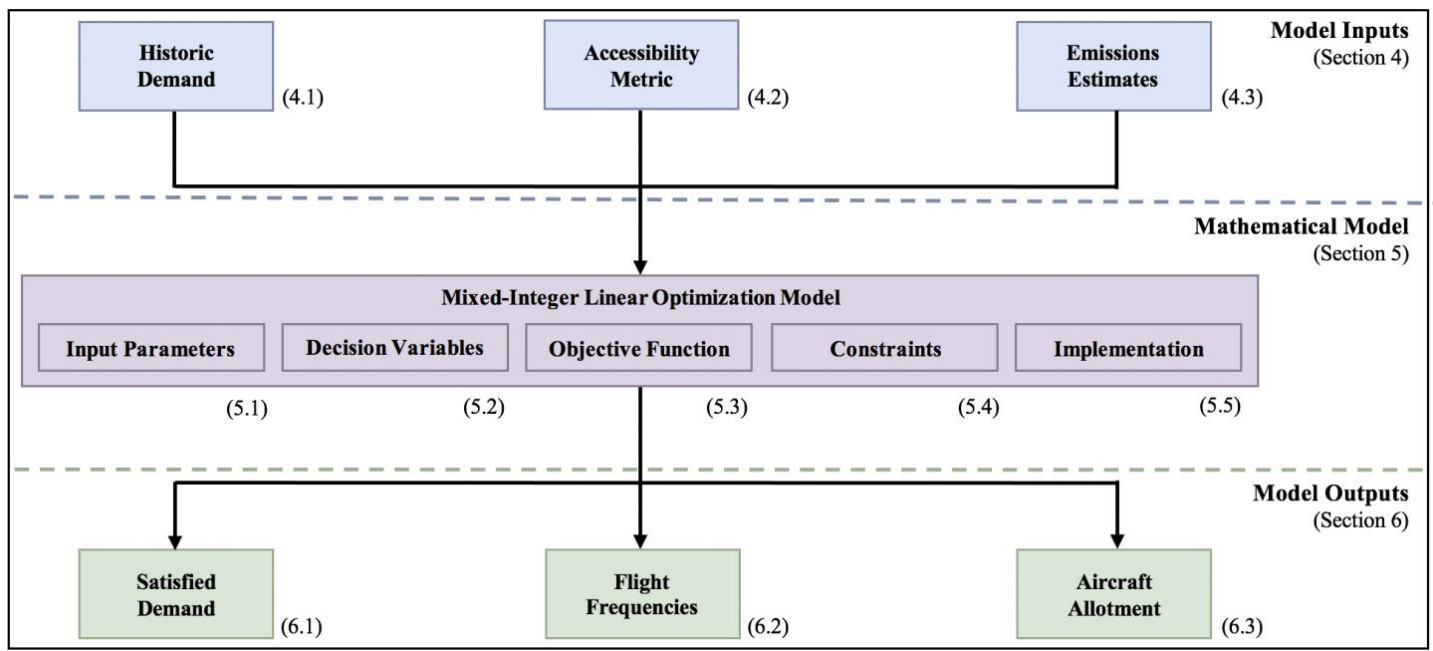

**Fig. 1.** Schematic of the model's main components. The numbers in parenthesis denote the section of the paper that explains the topic in question.

2-segment, one-way itinerary Philadelphia (PHL) → Houston (IAH) → Tampa (TPA) would count only for the historic demand between Philadelphia (PHL), the origin, and Tampa (TPA), the destination.

| United 2019 Q3 Itinerary Characteristics | | | |
|---|---|---|---|
| Unique Airports Visited | | 109 | Example Itinerary |
| Itineraries Sampled | | 528,974 | |
| % One-Way | | 41.67% | PHL (O) →IAH →TPA (D) |
| % Round-Trip | Even Legs | 51.38% | PHL (O) → ORD → MCI (D) →IAD → PHL (O) |
| | Odd Legs | 6.95% | PHL (O) → AUS (D) → EWR → PHL (O) |

**Table 1.** Statistics for the itineraries operated by United Airlines during 2019 Q3, excluding subsidiary flights operated through United Express. Overall, 14,895,200 passengers were transported during this period. In the example itinerary column, (O) denotes the origin airport, while (D) the destination.

Finding the historic OD demand for the round-trip itineraries was slightly more complicated and required introducing some statistical assumptions related to the total number of flight legs. For instance, in the case where the total number of flight legs in the itinerary was even (giving an odd total number of airports in the itinerary), the origin airport would be assumed to be the first in the sequence and then the destination airport to be the one falling in the middle of the sequence. As an example, a passenger flying on the 4-segment, round-trip itinerary Philadelphia (PHL) → Chicago (ORD) → Kansas City (MCI) → Washington DC (IAD) → Philadelphia (PHL) would only count for the historic OD demand between Philadelphia (PHL) and Kansas City (MCI) as well as in reverse between Kansas City (MCI) and Philadelphia (PHL). Notice that this approach inherently assumed symmetry in the historic OD demand for the round-trip itineraries.

However, in the case in which the total number of flight legs in the round-trip itinerary was odd (giving an even total number of airports in the itinerary), the origin airport would be taken, like before, as the first in the sequence, but then the destination airport as the one falling in the middle, taking the floor of the number of airports divided by two. In other words, a passenger flying on the 3-segment, roundtrip itinerary Philadelphia (PHL) → Austin (AUS) → Newark (EWR) → Philadelphia (PHL) would only count for the historic OD demand between Philadelphia (PHL) and Austin (AUS) as well as between Austin (AUS) and Philadelphia (PHL). To address the uncertainty on the OD pair demand estimation for odd-segmented round-trip itineraries under the utilized assumption, the present study also considered taking the ceiling of the number of airports divided by two. For instance, in the PHL → AUS → EWR → PHL example itinerary, PHL would still be the origin airport, but EWR the destination, as opposed to AUS before. Comparing the total historic demand for each of the possible 5,886 OD pairs in UA's network (complete graph of the 109 serviced airports) obtained from employing each assumption, it was found that both resulted in almost negligible differences. This outcome makes sense in the context of the small percentage (6.95%) that odd-segmented round-trip itineraries occupy over the total number of itineraries.

*4.2. Accessibility Metric*

Inspired by the work performed by Geurs and Van Wee (2004), Reynolds-Feighan and McLay (2006), Matisziw and Grubesic (2010) and Wei and Grubesic (2015), the present study proposes a new metric to measure airport accessibility, given by the ratio between the population served and the number of outbound enplanements (outbound passengers) at an airport. In essence, this metric compared the number of passengers that could potentially use an airport (population served), due to proximity (cost of access) and connectivity (quality of access), against those that actually did (outbound enplanements). Theoretically, in an equitably distributed system, all airports should have similar accessibility ratios. That is, someone, regardless of where it is living, has the same relative opportunity to not only access the air transportation system but also use it to reach its destination. However, in reality, due to competition, hubbing, and existing infrastructure (such as gate availability), certain airports are utilized much more extensively, resulting in a higher accessibility ratio (i.e., higher number of enplanements per inhabitant served). As a result, in the context of this metric, whether an airport is over or under-accessed will depend on its position with respect to the median accessibility ratio; over and under-accessed airports would fall above and below this median ratio, respectively.

Thus, to create this metric to quantify airport accessibility in the United States, first, the population served at each airport, $s_n$ ($n \in N$ denoting each airport) had to be calculated. In this work, $s_n$ was defined as a function of three main parameters: the distance from the airport to each census tract in the United States, $u_{n,t}$, the population of each census tract, $p_t$, and the importance of the airport, $\alpha_n$, where $t \in T$ denoted each census tract. A list of the variables composing the airport accessibility metric can be found on Table 2.

| Term | Definition |
|---|---|
| $N$ | Set of airports, indexed by $n$ |
| $T$ | Set of census tracts, indexed by $t$ |
| $I$ | Set of flight legs, indexed by $i$ |
| $I_n^{out} \subset I$ | Set of outbound flight legs from airport $n \in N$ |
| $s_n$ | Population served at airport $n$ |
| $u_{n,t}$ | Distance from airport $n$ to census tract $t$ |
| $p_t$ | Population of census tract $t$ |
| $f_i$ | Total frequency on flight leg $i$ |
| $M$ | Weighted directed adjacency matrix of the airline network |
| $\lambda_p$ | Dominant eigenvalue of $M_a$ |
| $v_p$ | Principal eigenvector of $M_a$ |
| $\alpha_n$ | Kleinberg (hub) centrality of airport $n$ |
| $\tau_t$ | Locational accessibility of each census tract $t$ |
| $\gamma_{t,n}$ | Catchment proportion of tract $t$ with respect to airport $n$ |
| $d_i$ | Historic OD demand of flight leg $i$ |
| $c'_n$ | Accessibility of airport $n$ |
| $c_n$ | Accessibility of airport $n$ (normalized) |

Table 2. Input parameters for the accessibility metric.

Here, $u_{n,t}$ was calculated as the haversine distance between the airport's coordinates as provided by the BTS's Master Coordinate dataset (BTS, 2023), and the census' tract centroid, as given by US Census data from the ACS's 1-year demographic and housing estimates (ACS, 2021). The importance of the airport, $\alpha_n$, was found through the Kleinberg (hub) centrality measure, which in the context of this work, required computing the principal eigenvector, $v_p$, of the weighted directed adjacency matrix of the airline network, $M$ (also known as the weighted connectivity matrix) multiplied by its transpose. More specifically, this calculation was given by:

$$MM^T v_p = \lambda_p v_p \qquad (1)$$

where $M$ was a square matrix. Specifically for UA, the dimension of this matrix was 109 (the same as the number of airports, or nodes in the graph representation of the route network), and its entries (the weights of the edges of the graph) were assigned through the following equation:

$$M_{(O,D)} = \begin{cases} f_i, & \text{flight existed between O and D} \\ 0, & O = D \\ 0, & \text{flight did not exist between O and D} \end{cases} \qquad (2)$$

where $f_i$ represented the total frequency (or number of flights) on leg $i$ between airports O (the origin) and D (the destination). Here, the variable $f_i$ inherently depended on the direction of the flight leg, given that the number of flights on one direction (e.g., O → D) need not necessarily be the same as in the other (e.g., D → O). $M$ was populated using the scheduling information from BTS's DB1B dataset (BTS, 2019) specific for 2019 Q3 and United Airlines. Overall, the matrix possessed 1,432 non-zero entries, which meant that UA operated a total of 716 unique routes, 12.2% of the 5,886 possible OD pairs in the network.

The vector $v_p$, also 109-dimensional, was found through its association with the eigenvalue $\lambda_p$, which had the largest absolute magnitude among all the eigenvalues. Then, the airport centralities, $\alpha_n$, could be extracted from $v_p$ as its individual elements, through:

$$v_p = (\alpha_1, \alpha_2, \alpha_3, \cdots, \alpha_{109}). \qquad (3)$$

These airport centralities provided a measure of the connectivity of a particular airport within the greater airline network, in terms of the direct flights that it possessed. Thus, the $\alpha_n$ could be used to augment the typically employed gravity model in transportation planning to construct a simple metric of locational accessibility for each census tract to the air transportation network, based on the basic potential interaction model introduced by Harris (1954):

$$\tau_t = \sum_{n \in N} \frac{\alpha_n}{u_{n,t}^2}, \qquad \forall t \in T \qquad (4)$$

Succinctly, this sum tried to capture both the cost of accessing the air transportation system from a census tract based on proximity ($u_{n,t}$) to a particular airport, and the access opportunities provided by that airport in terms of its importance and connectivity in the network ($\alpha_n$). This computation employed an inverse squared distance function, since this approach would prioritize airports geographically closer to the census tract in question. However, it should be noted that these results were airline dependent. Indeed, using Philadelphia County as an example, Newark Liberty (EWR) had the highest individual $\alpha_n/u_{n,t}^2$ access value, representing 66.8% as a fraction of $\tau_t$, as it was a major hub for United Airlines. La Guardia (LGA) followed with 21.0%, Philadelphia International (PHL) with 11.3%, and Harrisburg International (HIA) with 0.9%. Taking American (AA) instead of United would have yielded a distribution skewed towards PHL, since the airport is one of its east coast hubs.

This locational accessibility metric for census tracts was then used to create a catchment model for airports, defining a distribution of passengers to proximal airports, weighted by the airports' centralities. More specifically, to compute the expected population served at each airport, $s_n$, a sum was performed across all census tracts (of a total of 84,414), where the tract's population, $p_t$, was multiplied by the catchment proportion $\gamma_{t,n}$ for that census tract corresponding to a particular airport. Here, the former was obtained from the ACS's 1-year demographic and housing estimates (ACS, 2021). The latter, instead, was a computation which was hinted upon in the previous example, given by the fraction:

$$\gamma_{t,n} = \frac{\left(\frac{\alpha_n}{u_{n,t}^2}\right)}{\tau_t}, \qquad \forall t \in T, \forall n \in N. \qquad (5)$$

This value essentially denoted the expected percentage of a particular county's $t$ population that would choose to go to a particular airport $n$, based on distance (cost of access) and on that airport's centrality in the network (quality of access). In a similar manner to Fornito et al. (2016), this measure assumed that passengers would travel further in order to access larger, better connected airports due to expanded flight options, frequencies, and lower ticket fares. Overall, the sum describing $s_n$ resulted in eq. (6) below:

$$s_n = \sum_{t \in T} p_t \, \gamma_{t,n}, \qquad \forall n \in N. \tag{6}$$

Taking Philadelphia County's population once again as an example (still subject to the UA-specific data), it was found that the 11.3% of the population that would choose to go to PHL, as given by $\gamma_t$, would actually represent 46.1% of that airport's $s_n$ value. But the 66.8% that would instead opt for EWR would only signify 41.1% of that airport's $s_n$ (due to also capturing populations from the greater New York area). This information is summarized in Table 3 below. As this example illustrated, theoretically, all airports in the United States were candidates for the catchment of a particular county's population, although not all of them were necessarily suitable, due to the reasons listed above. Furthermore, as was mentioned earlier, these results were strictly airline dependent.

| Philadelphia County Population Distribution | | |
|---|---|---|
| Airport ($n$) | Catchment proportion ($\gamma_{t,n}$) | Percentage of $s_n$ |
| PHL | 11.3% | 46.1% |
| EWR | 66.8% | 41.1% |
| LGA | 21.0% | 11.8% |
| HIA | 0.9% | 1.0% |

**Table 3.** Population distribution of Philadelphia county across four airports employing UA-specific data. Here, although 21.0% of the county's population would be inclined to go to La Guardia as their airport of choice, that percentage would only represent 11.8% of the total population served by that airport.

Finally, to define the computed airport access ratio, $c'_n$, the total number of historic outbound enplanements (i.e., number of all departing passengers) was divided by the computed expected populations served at each airport, $s_n$, resulting in:

$$c'_n = \frac{\sum_{i \in I_n^{out}} d_i}{s_n}, \qquad \forall n \in N. \tag{7}$$

In eq. (7), the numerator was simply the addition of the historic OD demand $d_i$ of all flight legs $i$ out of the airport $n$. Finally, to ensure that the accessibility scores had a median of 1, the $c'_n$ values were divided by the median of their distribution to get the $c_n$ used in the optimization:

$$c_n = \frac{c'_n}{med(c'_n)}, \qquad \forall n \in N. \tag{8}$$

This framework facilitated computing the access scores of 109 airports in United Airlines' network as a case study, using 2019 Q3 passenger data and statistics from 84,414 census tracts in the United States (ACS, 2021), as shown in **Fig. 2**. Here, every census tract was colored following its respective locational accessibility value $\tau_t$, where the darker-colored areas had the best access to United's route system and the lighter-colored regions showed the least access. In addition, each airport was also identified with a circle proportional to its Kleinberg (hub) centrality $\alpha_n$; the present study's metric correctly determined Newark Liberty (EWR), Chicago O'Hare (ORD), Denver

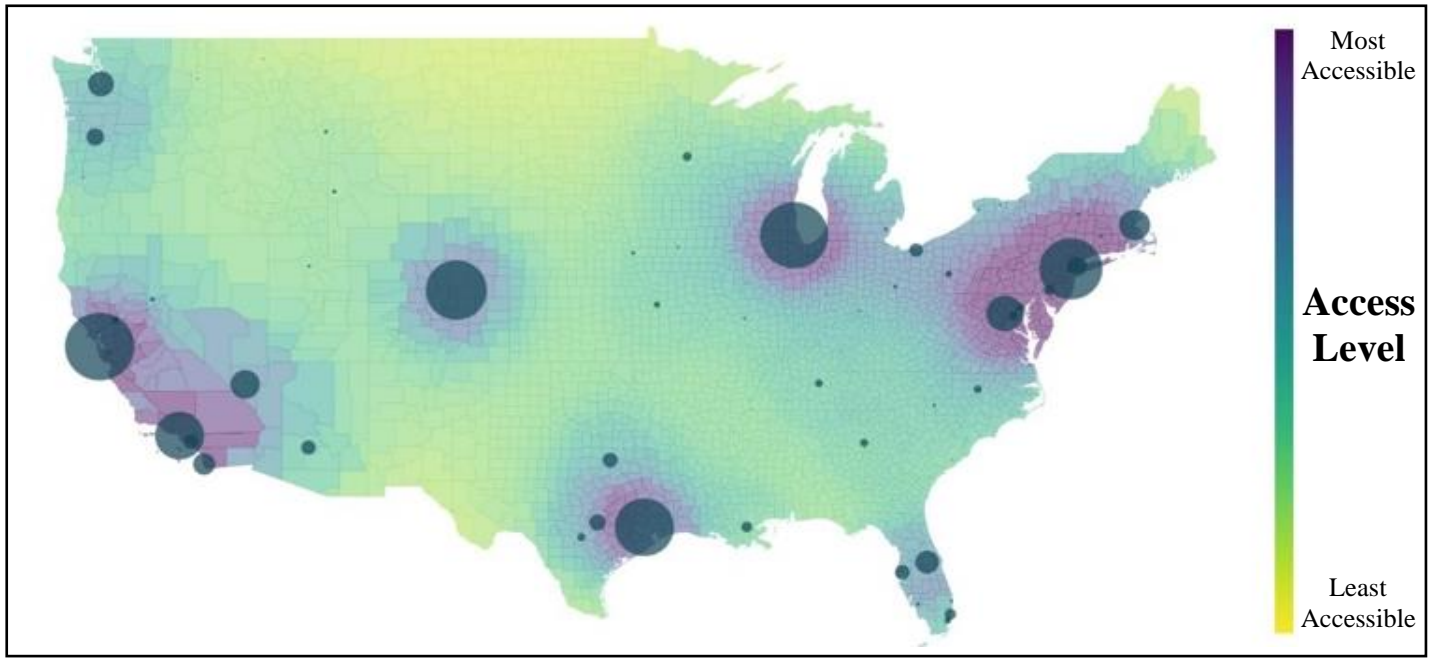

**Fig. 2.** United Airlines' passenger accessibility on the contiguous United States, for 2019 Q3 and 84,414 census tracts. The level of passenger accessibility (to United's route system) was illustrated using a color gradient, where more accessible parts of the country had darker-colored census tracts, while the lighter-colored regions showed the least access. The size of each circle, centered at an airport, was proportional to its Kleinberg (hub) centrality score.

International (DEN), San Francisco International (SFO) and Houston Intercontinental (IAH) as the most important airports for United, with Los Angeles International (LAX) and Washington Dulles (IAD) as secondary hubs.

### 4.3. Emissions Estimation Methodology

The procedure to compute the emissions from flight can be succinctly described as follows. In essence, for each OD pair in consideration throughout the optimization, the taxi-in, taxi-out and flight times were extracted from BTS' Airline On-Time Statistics (BTS, 2022) database. Details associated with the particular aircraft type operating the flight, including its N-number and seat capacity, were found using BTS's Schedule B-43 Inventory (BTS, 2021). The aircraft's engine type was sourced from both the FAA's Tail Registry (FAA, 2022), FAA's Engine Code Table (FAA, 2022) and the International Civil Aviation Organization's (ICAO) Engine Emissions Databank (ICAO, 2016), the latter of which has information regarding 816 different engine types. Aircraft and engine-specific emissions factors were then utilized to calculate emissions for the two stages of flight, Landing & Takeoff (LTO) and Cruise, Climb, & Descent (CCD), as standardized by the ICAO (Trozzi and Lauretis, 2016).

These emissions factors were extracted from the ICAO's Aircraft Engine Emissions Databank (ICAO, 2016) and EUROCONTROL's Base of Aircraft Data (EUROCONTROL, 2016), for LTO and CCD, respectively. The emissions estimated for each flight took into account four different types of greenhouse gases, $CO_2$, $CO$, $NO_x$ and $HC$, whose impact was converted to their equivalent effect in $CO_2$ terms following the conversion factors outlined by Brander and Davis (2012). Overall, **Fig. 3** provides a graphic overview of the emissions estimation workflow, alongside the different datasets utilized at each step. More detailed information can be found within Eskenazi et al. (2022).

## 5. MATHEMATICAL MODEL

### 5.1. Input Sets and Parameters

The optimization model developed in the present research built on the study conducted by Birolini et al. (2021), taking as input parameters the variables shown in Table 4. In this study, *A* referred to the set of aircraft and engine type pairings that the airline employed as their fleet during the study period. The aircraft type helped infer the seat capacity of a flight while the engine type enabled determining the emissions produced. To find the set *A*, the present work utilized both the FAA's tail registry (FAA, 2022) in conjunction with the BTS's Schedule B-43 Aircraft Inventory (BTS, 2021). In the case of United

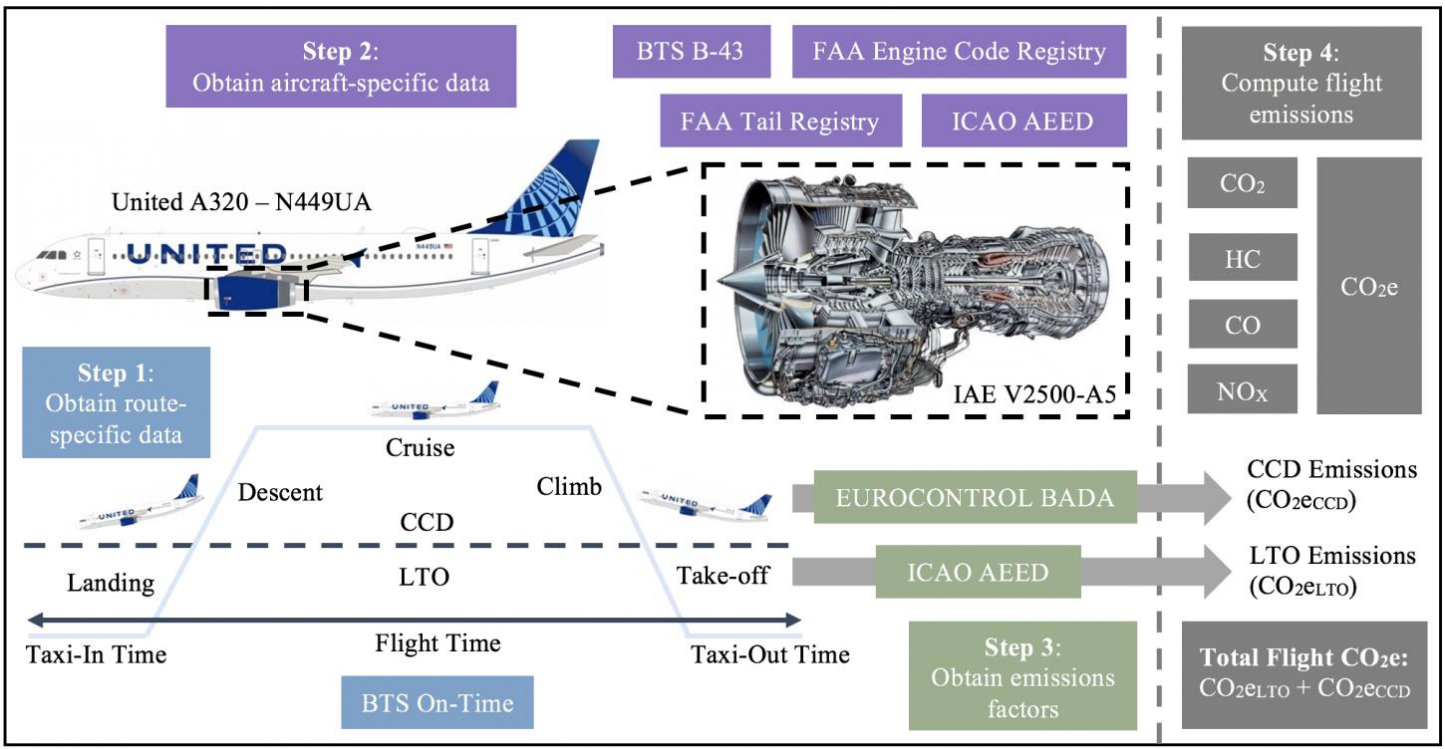

**Fig. 3.** Graphic workflow of the emissions estimation methodology utilized in the optimization. Each one of the colors (blue, violet, green, grey) represents a different step in the methodology. A total of six different open-source datasets are employed and integrated, including: BTS On-Time, BTS B-43, FAA Engine Code Registry, FAA Tail Registry, ICAO Aircraft Engine Emissions Dataset (AEED) and EUROCONTROL Base of Aircraft Data (BADA).

Airlines during 2019 Q3, the airline employed 66 different aircraft-engine pairings.

Next, the parameter *I* was defined as the set of flight legs *i* that the airline route network consisted of. This set was created by first constructing a graph with edges (flight legs) between the set of airports *N* where the historical demand (as calculated in Section 4.1) was greater than a pre-defined demand threshold. In the present study, a demand threshold of 100 daily passengers was used to construct the set *I*, equivalent to an 80% full A319 under UA's typical 126-seat configuration. The A319 is the smallest aircraft in United's mainline fleet (subsidiary United Express is excluded in the case study), so this threshold could guarantee at least one daily flight on this type on the considered OD pair. Then, the set of inbound flight legs to an airport was denoted by $I_n^{in}$, while the set of outbound flight legs from an airport was described by $I_n^{out}$.

The distance covered for each flight leg *i*, represented by the variable $y_i$, was taken as the haversine distance between the origin and destination airports. The computation of $y_i$ required using coordinate information sourced from the BTS's Master Coordinate dataset (BTS, 2023).

Continuing, $A_i^0$ was the set of aircraft and engine type pairings that could not be used to operate a flight on a particular leg *i*. This set was constructed by using specific aircraft type ranges, extracted from Palt (2019), as a threshold. In essence, if the aircraft's range was less than the distance to be flown on a flight leg, then the aircraft type could not be used to operate a flight on that route.

Next, the parameter $k_a$ represented the seat capacity for a particular aircraft type. This variable, mainly sourced from BTS's Schedule B-43 Inventory (BTS, 2021), helped inform the possible number of passengers that could be accommodated on a particular flight. To reduce the level of complexity in the optimization, the present study did not distinguish between classes of service (e.g., economy or business). Hence, for the purposes of passenger allocation on flights and computations of $CO_2$e/seat-mile, the underlying assumption was that all passengers were flying in economy.

Another variable, $l_a$, denoted the maximum utilization time of an aircraft. This parameter took into account the fact that an aircraft could only fly for a limited amount of time in a given day due to maintenance and turnaround times. The procedure to determine the maximum utilization time for an aircraft type simply involved finding the time during the study period in which the cumulative number of hours in the air was maximized at a particular day. BTS's On-Time Performance (BTS, 2022) and Schedule B-43 Inventory (BTS, 2021) datasets facilitated this procedure, since the former contained data for every flight that included airtime and an associated N-number that operated it, while the latter could be used to find the aircraft type from the N-number.

A related quantity, $t_{i,a}$, was defined as the block time for a given flight leg and aircraft. In this work, the block time for a flight was given by adding the expected taxi-out and taxi-in times and the airtime. Similarly to $l_a$, both BTS's On-Time Performance (BTS, 2022) and Schedule B-43 Inventory (BTS, 2021) datasets were employed to compute $t_{i,a}$.

The fleet size for an airline was captured using the parameter $w_a$, which denoted the number of aircraft of a specific model possessing a certain engine type (in other words, the aircraft-engine pairings). This variable was determined from using both the FAA's tail registry (FAA, 2022) and BTS's Schedule B-43 Aircraft Inventory (BTS, 2021).

The parameter $c_n$ represented the accessibility score of an airport, as determined through the procedure described above in Section 4.2. In parallel, $e_{i,a}$ denoted the emissions associated with a flight leg *i*, for a particular aircraft and engine type pairing *a*. These emissions were calculated using the estimation methodology outlined in Section 4.3., as developed by the authors.

| Term | Definition |
|---|---|
| $N$ | Set of airports, indexed by *n* |
| $A$ | Set of aircraft-engine pairings, indexed by *a* |
| $A_i^0 \subset A$ | Subset of aircraft-engine pairings that cannot operate leg *i* |
| $I$ | Set of flight legs, indexed by *i* |
| $y_i$ | Distance of flight leg *i* |
| $I_n^{in} \subset I$ | Set of inbound flight legs to airport $n \in N$ |
| $I_n^{out} \subset I$ | Set of outbound flight legs from airport $n \in N$ |
| $k_a$ | Seat capacity of aircraft-engine pairing *a* |
| $l_a$ | Maximum utilization of aircraft-engine pairing *a* |
| $t_{i,a}$ | Block time of leg *i* operated with aircraft-engine pairing *a* |
| $w_a$ | Number of aircraft-engine pairings of type *a* |
| $c_n$ | Accessibility of an airport *n* |
| $d_i$ | Historic OD demand of flight leg *i* |
| $e_{i,a}$ | Emissions of leg *i* operated with aircraft-engine pairing *a* |

**Table 4.** Input Parameters for the Mixed-Integer Linear Optimization Model.

*5.2. Decision Variables*

The decision variables used in the mathematical model (i.e., the parameters whose values would be determined through the optimization) can be seen in Table 5. Consisting of three variables, the first, $f_{i,a}$, represented the frequency of flights operated on a given flight leg *i* using a particular aircraft and engine type pairing *a*. The second, $f_i$, denoted the total frequency, across all aircraft and engine type pairings, for a particular flight leg *i*. Finally, $q_i$, captured the total number of passengers accommodated on each flight leg *i*.

| Term | Definition |
|---|---|
| $f_{i,a} \in \mathbb{Z}^+$ | Frequency on flight leg *i* operated by aircraft-engine *a* |
| $f_i \in \mathbb{Z}^+$ | Total frequency on flight leg $i$ |
| $q_i \in \mathbb{Z}^+$ | Number of passengers accommodated on flight leg $i$ |

**Table 5.** Decision Variables for the Mixed-Integer Linear Optimization Model. In this table, $\mathbb{Z}^+$ denoted the set of non-negative integers, i.e., 0, 1, 2...

*5.3. Objective Function*

The objective function of the present study can be understood as the mathematical representation of the trade-off between emissions and passenger accessibility. In essence, this expression, given by eq. (9) below, aimed to maximize the economic benefit associated with flying passengers in the route network while reducing the cost emissions per mile flown.

$$max \quad V \sum_{i \in I} \frac{q_i}{y_i} - G \sum_{i \in I} \sum_{a \in A} f_{i,a} e_{i,a} \quad (9)$$

The weights of the two main components of the function were given by the parameters $V$ and $G$. As shown on Table 6 below, the variable $V$ indicated the monetary value associated with each additional passenger per mile flown while $G$ symbolized the social cost of $CO_2e$ per emitted ton. Evidently, because of the nature of the weighting parameters, the objective function was purely economic in nature, since it provided a framework that enabled comparing emissions and passenger accessibility on the same units.

| Term | Definition |
|---|---|
| $V$ | Monetary Value provided by each passenger per mile flown |
| $G$ | Social Cost of $CO_2e$ per ton |

**Table 6.** Economic Weights for the Objective Function.

*5.4. Constraints*

The constraints behind the optimization model, shown in equations (10) through (15), sought to replicate the complexities of route network planning and as a result drove the model to output realistic results. To start, eq. (10) ensured that the number of inbound flights of a particular aircraft and engine type pairing at an airport was equal to the number of outbound flights with the same aircraft and engine type pairing. Next, eq. (11) guaranteed that the time for which an aircraft was operative during a day was less than or equal to its maximum utilization time. Eq. (12) enforced that an aircraft could not be deployed to operate a flight leg whenever its range was less than the distance to be travelled on the flight leg. On similar lines, eq. (13) established that the number of passengers flying did not exceed the number of seats available. Then, eq. (14) dictated that the sum of flight frequencies across different aircraft and engine pairings should be equal to the total flight frequency on a flight leg. Finally, and perhaps most importantly, eq. (15) warranted that airports with an existing high accessibility score did not serve a larger number of passengers than they previously did and that airports with low accessibility scores served at least as many passengers as they previously did, if not more.

$$\sum_{i \in I_n^{in}} f_{i,a} = \sum_{i \in I_n^{out}} f_{i,a}, \quad \forall n \in N, \forall a \in A \quad (10)$$

$$\sum_{i \in I} f_{i,a} t_{i,a} \leq l_a w_a, \quad \forall a \in A \quad (11)$$

$$f_{i,a} = 0, \quad \forall a \in A_i^0, i \in I \quad (12)$$

$$q_i \leq \sum_{a \in A} f_{i,a} k_a, \quad \forall i \in I \quad (13)$$

$$f_i = \sum_{a \in A} f_{i,a}, \quad \forall i \in I \quad (14)$$

$$\frac{d_i}{\max(c_n, 1)} \leq q_i \leq \frac{d_i}{\min(c_n, 1)}, \quad \forall i \in I_n^{out}, \forall n \in N \quad (15)$$

*5.5. Implementation*

The present study implemented the optimization model on IBM's CPLEX MILP Solver (v12.9), utilizing United Airlines route network data during the months of July, August, and September of 2019 as a case study. The optimization was performed a total of five times, enforcing a maximum runtime of 10 hours, after which an approximat optima, found by that time, would be returned. Overall, the average number of decision variables across the five optimizations exceeded 200,000 while the average number of constraints surpassed 100,000. Naturally, because of the scale and computational complexity of the problem being solved due to the integer decision variables, finding the global maxima would require significantly more runtime. However, the present study found that the system-wide improvements achieved by the optimization, when measured across the four system-wide categories (as shown in Table 6) and allowed a runtime of 20 hours, did not differ notably from those results obtained with the 10-hour runtimes.

In all five cases, the weighs of the two components guiding the objective function in eq. (9) were selected using governmental data from the United States Department of Transportation. More specifically, $V$, the economic value associated with each additional passenger per mile flown, was set to $50-mile/passenger, as this was in the range of how the United States' government priced and valued EAS flights (U.S. Department of Transportation, 2022). For $G$, the social cost of emitted ton of $CO_2e$, the chosen monetary value was $50/ton, since this number was in agreement with that suggested by the Biden administration (White House, 2021).

## 6. RESULTS

Although the approximate optima route networks returned by each of the five optimizations differed in terms of the specific selected OD pairs, aircraft flown on each segment or system-wide improvements, when seen from a global perspective, each succeeded in balancing the trade-off between emissions and passenger accessibility. A sample optimized route network for United Airlines is shown in **Fig. 4**, while the average system-wide improvements achieved by the five optimizations are shown on Table 7, in the "Optimized Network" row. As a benchmark for comparison, "2019 Q3 Network" refers to the system-wide statistics corresponding to UA's route network during the case study period.

| Term | Number of Flights | Demand Capacity | $CO_2$e/seat-mile | Accessibility Scores |
|---|---|---|---|---|
| 2019 Q3 Network | 164,350 | 14,895,200 | 0.384 | - |
| Optimized Network | 127,426 | 16,799,641 | 0.367 | - |
| Percentage Change | 22.4% reduction | 12.8% increase | 4.4% reduction | 17.6% SD reduction |

**Table 7.** System-wide average improvements for United Airlines Q3 2019 route network.

Across all four of the reported metrics (number of flights, demand capacity, $CO_2$e/seat-mile and accessibility scores), the optimized networks showed enhanced performance compared to the original 2019 Q3 network. However, this result is strictly dependent on the set of weights of the objective function, *V* and *G*, which in the present study were set to $50-mile/passenger and $50/ton, respectively.

### 6.1. Satisfied Demand

The optimized route network suggested that United should have operated 127,426 flights over the course of 2019 Q3, a 22.4% reduction compared to the 164,350 flights the airline actually flew. Despite flying less, in reality, the model determined that there was enough capacity to support the demand of 16,799,641 passengers. Interestingly, during the case study period, UA offered an estimated total of 14,895,200 seats. Thus, under optimized conditions, it could have been possible to increase the number of seats offered by 12.8%. These stark disparities, between the actual 2019 and optimized network performances, could likely be the result of competition for infrastructure at airports, complex supply-demand interactions, and varying load factors, all of which are not currently addressed in this model.

### 6.2. Flight Frequencies

The model also suggested that to best reduce aggregate emissions and improve passenger accessibility, significant modifications on flight frequencies would need to be performed. For the former, the optimization managed to achieve reductions in the system-wide $CO_2$e/seat-mile ratio by favoring long-haul, direct flights over short-haul, connecting ones, at least whenever OD passenger demand was sufficient to sustain those. Examples of new, direct long-haul flights introduced by the model in UA's route network included Las Vegas (LAS) to Atlanta (ATL), Charleston (CHS) to Los Angeles (LAX), and Billings (BIL) to Dallas Fort Worth (DFW). Overall, these new, direct, longer flights allowed the model to support an average system-wide $CO_2$e/seat-mile of 0.367, compared to UA's 2019 Q3 historic value of 0.384 (a 4.4% reduction). This overall tendency was in agreement with the results from the previously performed work by Peeters et al. (2001), who found that increasing the number of longer point-to-point routes as opposed to shorter hub-and-spoke routes reduced an airline network's cumulative emissions.

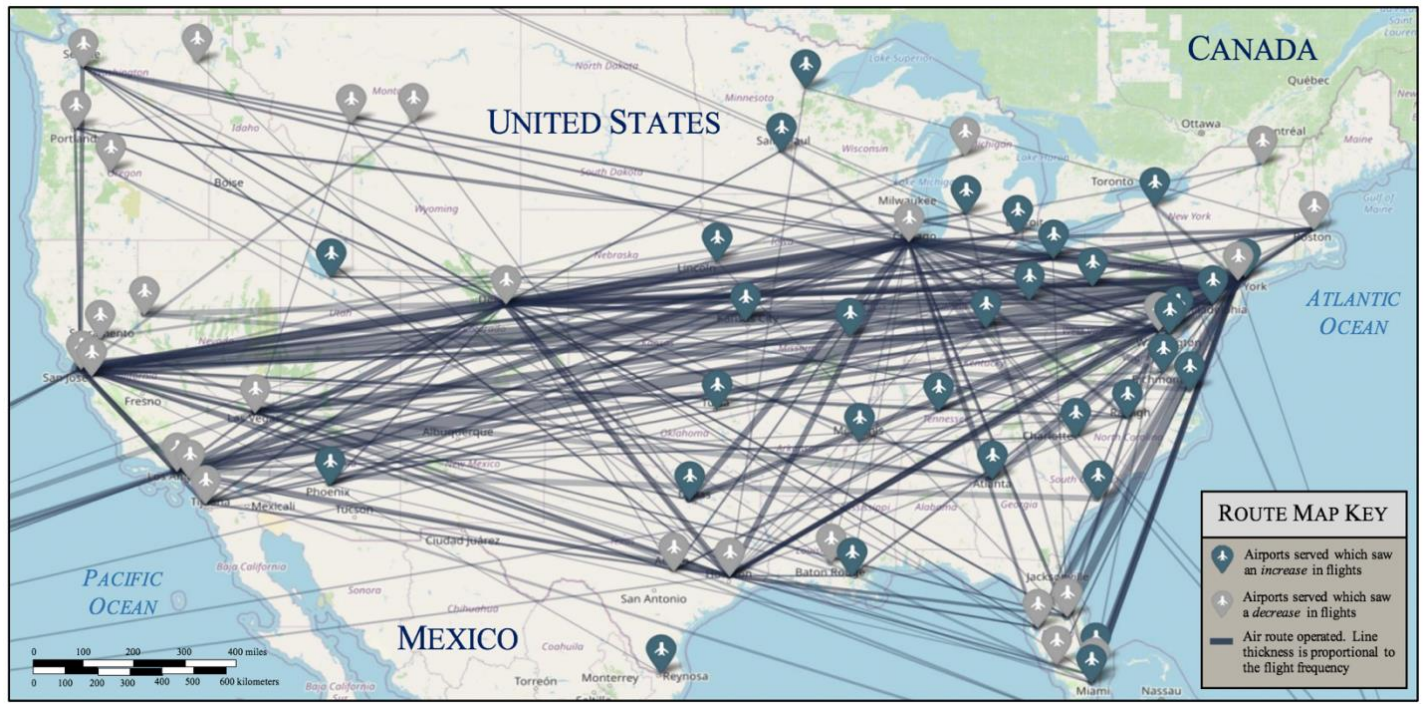

**Fig. 4.** United Airlines' optimized route networks for 2019 Q3. This plot only displays the flight legs for which there were at least one daily flight during the study period. Airports which saw an increase in outbound enplanements were identified with a blue pin, while airports whose service had decreased were colored with a gray pin. The thickness of a line connecting two airports is proportional to the assigned flight frequency along that leg.

For the latter, the model succeeded in distributing service more equitably (i.e., reducing the spread of the airport accessibility scores) by shifting operations from hubs to direct point-to-point routes. Indeed, as **Fig. 4** evidences, virtually all of UA's hub airports (IAD, EWR, ORD, DEN, SFO, LAX, and IAH) saw a reduction in the amount of service they received, as identified by the grey pin. In contrast, the vast majority of airports in the optimized network, as identified with a blue pin, instead saw increases in air service, with Duluth (DLH), Rochester (ROC), Memphis (MEM), Tulsa (TUL) and Lincoln (LNK) experiencing the highest percentage improvements in their accessibility scores. Examples of routes proposed by the optimization involving these airports included Duluth to Boston (BOS), Lincoln to Orlando (MCO), and Memphis to Baton Rouge (BTR), all three of which are currently not operated by any commercial airline in the United States.

The logic behind service reductions at hub airports can be explained with the fact that these cities possessed artificially high accessibility scores (i.e., a disproportional number of enplanements per population served) due to their status as network hubs. As an example, both Denver and Memphis are similarly sized cities (ACS, 2021) with similar catchment populations, yet DEN had a significantly higher accessibility score than MEM. Thus, cutting service from DEN (a UA hub) to make aircraft available to open new, non-stop routes between lower accessibility airports (such as between MEM and BTR) would have distinct implications for different passengers in the network. More precisely, for the typical DEN passenger, the service reduction would not dimmish its flight options due to the hub still possessing a high level of connectivity and flight frequencies. But for the passenger flying out of the less accessible MEM, the opening-up of service would have a tremendous impact, since percentage wise, it would affect that airport's accessibility score more than it would to the hub.

Overall, through the optimization's re-arrangement of flights, service was indeed more equitable across all airports, as measured by the metric developed in section 4.2. However, instead of reporting the specific accessibility scores of each airport, the present study focused on examining the spread of these, i.e., the standard deviation of their distribution, or distance from the mean. Mathematically, the notion of an equitably serviced airline network translated into a distribution of airport accessibility scores that were bundled closer towards the mean. Compared to the historic distribution of 2019 Q3, the optimization succeeded in reducing by 17.6% the standard deviation of the accessibility scores. In other words, this meant that across all airports, the ratio of population served to number of outbound enplanements was more similar.

However, these observations and results are specific to the airline data utilized in the optimization, in particular that employed to estimate the historic OD demand (see section 4.1). Thus, caution must be exercised when extending conclusions to the entire airline industry in the United States. As an example, as featured in **Fig. 4**, the model returned that UA should have increased operations at Atlanta (ATL), Salt Lake City (SLC), Charlotte (CLT) and Phoenix (PHX) in order to distribute its airline service more equitably. However, in reality, the first two airports are currently hubs for Delta Airlines, while the latter two for American Airlines, and as a result, have no shortage of air service nor connectivity. Hence, to truly provide policy recommendations that would effectively help improve passenger accessibility in the United States, it would be necessary to overlay the airline data of all major carriers, eliminating the possibility of reaching misleading conclusions as in the preceding example. Nonetheless, this proposal, which could be explored in future work, would abruptly increase the complexity of the optimization to +1,000,000 decision variables and constraints, rendering the problem unsolvable in a reasonable timescale.

*6.3. Aircraft Allotment*

In addition to determining new flight frequencies along each leg while satisfying the historic demand, the model optimally allocated the airline's aircraft fleet. For instance, for flights between EWR and LAX, the aircraft and engines pairings were assigned according to Table 8.

To satisfy the 1,043 historic flights from EWR to LAX, United chose to use 11 different aircraft-engine pairings (Table 7 only shows 4 of these, 3 of which were different configurations of the same aircraft, the B757-200). Optimizing for reduced emissions, the model determined that United only needed to fly 800 times with just 3 aircraft-engine pairings to satisfy the same historic demand, even assuming a typical industry load factor of 85% (BTS, 2016).

| Route | Historic | | Optimized | |
|---|---|---|---|---|
| Flights | 1,043 | | 800 | |
| Aircraft Engine Usage | Boeing 757-200 RR RB.211 | 411 | Boeing 767-300 P&W PW4000 | 433 |
| | Boeing 757-200 P&W PW2037 | 155 | Boeing 737-900 CFM CFM56-7B27E | 361 |
| | Boeing 757-200 P&W PW2040 | 132 | Boeing 737-800 CFM CFM56-7B26/3 | 6 |
| | Boeing 787-900 GE GENX-1B76/P2 | 111 | | |
| | ... | ... | | |

**Table 8.** Differences in aircraft allotment and flights on the Newark Liberty International Airport (EWR) to Los Angeles International Airport (LAX) segment. Here, RR = Rolls-Royce, P&W = Pratt & Whitney, GE = General Electric, and CFM = CFM International. The reported total number of flights corresponded to those operated during the entire case study period, 2019 Q3. For a roughly 90-day quarter, 1,043 and 800 flights translated into approximately 12 and 9 flights per day, respectively.

## 7. CONCLUSIONS

The present study introduced a novel, open-source analytical framework and optimization model to restructure airline route networks in the United States to jointly minimize emissions and maximize accessibility. To achieve this task, the optimization model utilized three key inputs. First, it leveraged accessibility scores for 109 airports in the United States, which were obtained by utilizing a new strategy for measuring the accessibility of an airline route network, building on the previous work of Geurs and Van Wee (2004), Matisziw and Grubesic (2010) and Wei and Grubesic (2015). Second, it harnessed system-wide emissions estimations for +100,000 flights, which were calculated by incorporating granularities as the time spent in various stages of flights and the type of aircraft and engine used, building on the work of Peeters (2001) and Peeters and Williams (2012). Third, it used historic OD demand for 5,886 potential flight legs, which were extracted from +500,000 historic itineraries sampled.

Overall, these three inputs – accessibility, emissions and historic OD demand – facilitated the creation of an optimization model that successfully demonstrated that it is possible to jointly minimize flight emissions and maximize accessibility, rather than strictly maximizing profit. Bound by a rich set of +100,000 realistic constraints and +200,000 decision variables, the optimization returned a more efficient, restructured route network with less flying, equitably distributed flight frequencies that satisfied historic demand, and a simplified fleet allocation along each route that minimized the airline's environmental footprint. Using United Airlines as a case study during the 2019 Q3 period, the results indeed demonstrated that the airline could have decreased the average emissions per seat-mile by 4.4% and reduced the standard deviation of the airports' accessibility scores by 17.6%, all while satisfying passenger OD demand.

Future work could seek to implement this optimization framework on each of the major U.S. airlines, and further explore the trade-off between accessibility and emissions across them, as well as the effects of competition. Moreover, as was hinted earlier in section 6.2, overlaying the airline data of all major carriers to simulate a massive, single U.S. airline could also be another potential direction for future studies, given the valuable insights that it could provide about the United States' airline industry. As an example, this simulation could help inform airport infrastructure expansion decisions, through scenarios where, for instance, more flying takes place at an airport with limited gate capacity (or remote stand positions). Indeed, incorporating gate capacity as an additional constraint (like those in section 5.4) would make the overall optimization even more realistic. In addition, a sensitivity analysis that considers different values of $V$ (monetary value provided by each passenger per mile flown) and $G$ (social cost of $CO_2$e per ton) would be extremely useful to both policymakers and airlines; effectively, this analysis could help evaluate the effect of various carbon taxes on the aviation sector while still emphasizing the need to service under-accessed communities. On a similar note, examining the frontier of optimal Pareto solutions, using $V$ and $G$ as weights, would certainly be appropriate, although the complexity, magnitude and runtime of each optimization would demand a number of computational resources that the authors currently do not possess. Furthermore, in order to better reflect the realities of the aviation industry, the optimization model must be extended to more accurately represent load factors, connecting passengers, and supply-demand interactions. Finally, while the model as was presented in this paper only focused on network restructuring from a purely environmental-accessibility standpoint, future work could seek to incorporate flight profit and examine the economic costs of shifting from an existing route network to one optimized for reduced emissions and improved airport accessibility.

While much work remains to be done, the present study will hopefully help pave the way for more open-source, data-driven sustainable aviation research for policymaking. In the future, it will serve as a useful platform to simulate various costs of $CO_2$ emissions and economic benefits of having more passengers flying, so as to model scenarios that can motivate policy and network planning decisions to equitably fight climate change.

**AUTHOR COMPETING INTEREST**

All four authors contributed equally to the research behind this study and the writing of this manuscript.

**DECLARATION OF COMPETING INTEREST**

The authors declare that they have no known competing financial interests or personal relationships which have, or could be perceived to have, influenced the work reported in this article.## REFERENCES

Airbus. (2020). Airbus Reveals New Zero-Emission Concept Aircraft. https://www.airbus.com/en/newsroom/press-releases/2020-09-airbus-reveals-new-zero-emission-concept-aircraft

American Community Service (ACS). (2021). ACS Demographic and Housing Estimates. https://data.census.gov/table?tid=ACSDP1Y2021.DP05

Arul, S. G. (2014). Methodologies to monetize the variations in load factor and GHG emissions per passenger-mile of airlines. *Transportation Research Part D: Transport and Environment*, *32*, 411-420.

Birolini, S., Jacquillat, A., Cattaneo, M., & Pais Antunes, A. (2021). Airline Network Planning: Mixed-integer non-convex optimization with demand–supply interactions. *Transportation Research Part B: Methodological* 154, pp. 100–124. ISSN: 0191-2615. DOI: https://doi.org/10.1016/j.trb.2021.09.003. article/pii/S0191261521001703.

Brander, M., & Davis, G. (2012). Greenhouse gases, CO2, CO2e, and carbon: What do all these terms mean. Econometrica, White Papers.

Bureau of Transportation Statistics (BTS). (2016). Domestic Load Factor on U.S. Airlines, Seasonally Adjusted. https://www.bts.gov/content/table-2-domestic-load-factor-us-airlines-seasonally-adjusted-2

Bureau of Transportation Statistics (BTS). (2019). Airline Origin and Destination Survey (DB1B) Database. https://www.transtats.bts.gov/DatabaseInfo.asp?QO_VQ=EFI&DB_URL=

Bureau of Transportation Statistics (BTS). (2021). Schedule B-43 Inventory. https://rosap.ntl.bts.gov/view/dot/6438

Bureau of Transportation Statistics (BTS). (2022). Airline On-Time Statistics. https://www.transtats.bts.gov/ontime/

Bureau of Transportation Statistics (BTS). (2023). Master Coordinate. https://www.transtats.bts.gov/tableinfo.asp

Eskenazi, A., Butler, L., Joshi, A., & Ryerson, M. (2022) Democratizing Aviation Emissions Estimation: Development of an Open-Source, Data-Driven Methodology. *10th International Conference on Research in Air Transportation.* DOI: 10.48550/ARXIV.2202.11208.

ETC. (2021). Ten Critical Insights on the Path to a Net-Zero Aviation Sector". *Energy Transitions Commission*.

EUROCONTROL. (2013). User manual for the base of aircraft data (BADA) revison 3.11." Brussels: Eurocontrol.

Fageda, X., Suárez-Alemán, A., Serebrisky, T., & Fioravanti, R. (2018). Air connectivity in remote regions: A comprehensive review of existing transport policies worldwide. Journal of Air Transport Management, 66, 65-75.

Federal Aviation Administration (FAA). (2022). Tail Registry. https://www.faa.gov/licenses_certificates/aircraft_certification/aircraft_registry/releasable_aircraft_download

Filippone, A., & Parkes, B. (2021). Evaluation of commuter airplane emissions: A European case study. *Transportation Research Part D: Transport and Environment*, *98*, 102979.

Fornito, A., Zalesky, A., & Bullmore, E. T. (2016). Chapter 5 - Centrality and Hubs. Fundamentals of Brain Network Analysis. *Academic Press*, pp. 137–161. ISBN: 978-0-12407908-3. DOI: https://doi.org/10.1016/B978-0-12407908-3.00005-4.

Geurs, K. T., & Van Wee, B. (2004). Accessibility evaluation of land-use and transport strategies: review and research directions. *Journal of Transport geography* 12.2, pp. 127–140.

Graver, Brandon, Kevin Zhang, and Dan Rutherford. (2019). Emissions from commercial aviation, 2018. International Council on Clean Transportation.

Grubesic, T. H., Matisziw, T. C., & Murray, A. T. (2012). Assessing geographic coverage of the essential air service program. *Socio-Economic Planning Sciences* 46.2, pp. 124–135.

Grubesic, T. H., & Wei, F. (2013). Essential Air Service: a local, geographic market perspective. *Journal of Transport Geography* 30, pp. 17–25.

Grubesic T. H., & Wei., F. (2012). Evaluating the efficiency of the Essential Air Service program in the United States.
14